\documentclass[pdftex,superscriptaddress,floats,showpacs]{revtex4}
\usepackage{graphicx}
\usepackage{dcolumn}
\usepackage{bm}
\usepackage{amssymb}
\usepackage{amsmath}
\usepackage{mathrsfs}

\begin{document}

\title{Boosted high harmonics pulse from a double-sided relativistic mirror}
\author{T. Zh. Esirkepov}
\author{S. V. Bulanov}
\altaffiliation[Also at ]{A. M. Prokhorov Institute of General Physics of RAS, Moscow, Russia.}
\author{M. Kando}
\author{A. S. Pirozhkov}
\affiliation{Kansai Photon Science Institute, JAEA, Kizugawa, Kyoto 619-0215, Japan}
\author{A. G. Zhidkov}
\affiliation{Central Research Institute of Electric Power Industry, Yokosuka, Kanagawa
240-0196, Japan}
\date{January, 2009; rev. March, 2009.}

\begin{abstract}
An ultra-bright high-intensity X- and gamma-radiation source is proposed. A
high-density thin plasma slab, accelerating in the radiation pressure
dominant regime by a co-propagating ultra-intense electromagnetic wave,
reflects a counter-propagating relativistically strong electromagnetic wave,
producing strongly time-compressed and intensified radiation due to the
double Doppler effect. The reflected light contains relativistic harmonics
generated at the plasma slab, all upshifted with the same factor as the
fundamental mode of the incident light.
\end{abstract}


\maketitle


Interaction of electromagnetic (EM) wave with the relativistic mirror has
been used by A. Einstein to illustrate basic effects of special relativity 
\cite{bib:Einstein}. In modern theoretical physics the consept of
relativistic mirror is used for solving a wide range of problems,
such as the dynamical Casimir effect \cite{bib:Cas}, the Unruh radiation \cite
{bib:Unruh} and other nonlinear vacuum phenomena.
Relativistic mirrors made
by wake waves may lead to an electromagnetic wave intensification \cite
{bib:FM} resulting in an increase of pulse power up to the level when the
electric field of the wave reaches the Schwinger limit when electron-positron
pairs are created from the vacuum and the vacuum refractive index becomes
nonlinearly dependent on the electromagnetic field strength \cite{bib:NQED}.
In classical electrodynamics EM wave reflected off a moving mirror undergoes
the frequency and electric field magnitude multiplication, a phenomenon
called the double Doppler effect. If the EM wave is co-propagating with
respect to the mirror, its frequency and energy decreases upon reflection.
If it is counter-propagating, the reflected light gains energy and becomes
frequency-upshifted. For the latter case the multiplication factor is
approximately $4\gamma ^{2}$, where $\gamma \gg 1$ is the Lorentz factor of
the mirror, making this effect an attractive basis for a source of powerful
high-frequency EM radiation. Relativistic plasma provides numerous examples
of moving mirrors which can acquire energy from co-propagating EM waves or
transfer energy to reflected counter-propagating EM waves (see \cite{bib:Rev}
and references therein).
Manifestation of the Doppler effect in
plasma governed by strong collective fields is seen in the concepts of the
sliding mirror \cite{bib:FOIL},
oscillating mirror \cite{bib:OscM},
flying mirror \cite{bib:FM} and other schemes \cite{bib:ATTO} which can
produce ultra-short pulses of XUV- and X-radiation.

\begin{figure}[Hb]
\includegraphics[width=3.3in]{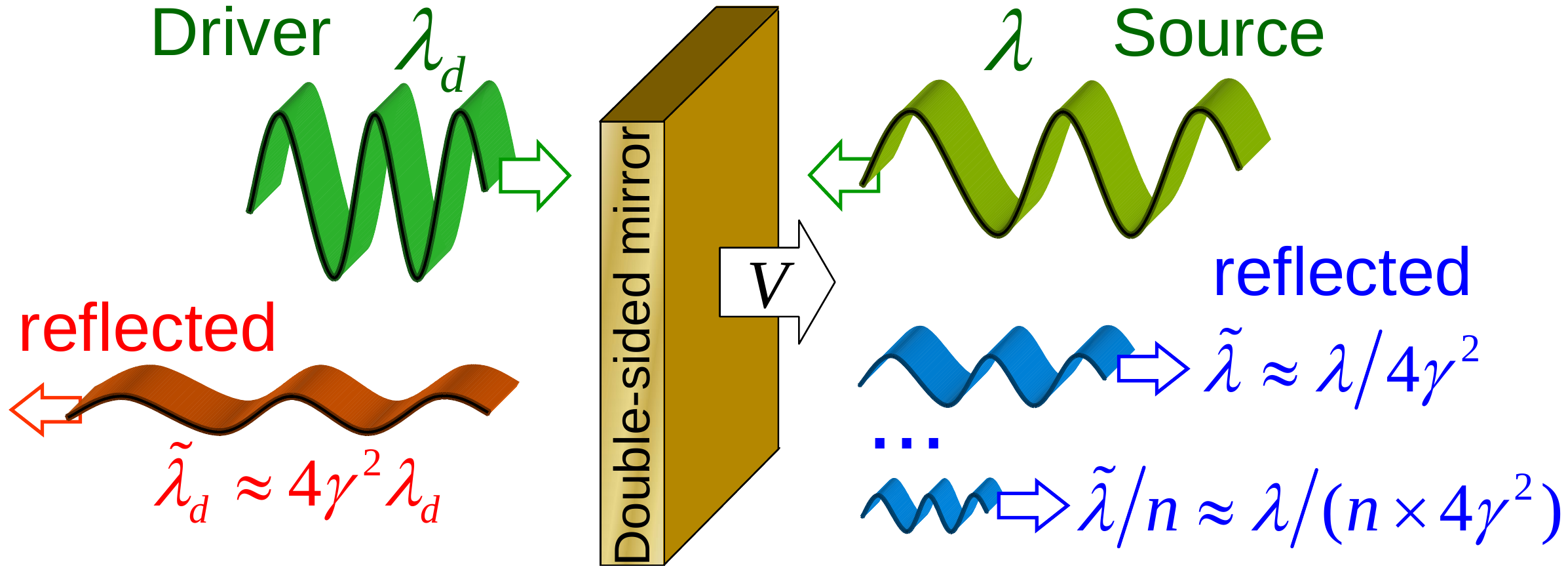}
\caption{(color)
The accelerating double-sided mirror concept. }
\label{fig:Kagami}
\end{figure}
In this paper we discuss the concept of the accelerating double-sided
mirror, Fig. \ref{fig:Kagami}, which can efficiently reflect the
counter-propagating relativistically strong electromagnetic radiation. The
role of the mirror is played by a high-density plasma slab accelerated by an
ultra-intense laser pulse (the driver) in the Radiation Pressure Dominant
(RPD) regime (synonymous to the Laser Piston regime), \cite{bib:RPD}. Such
an acceleration can be viewed as the double Doppler effect: it is the
reflection that allows the energy transfer from the driver pulse to the
co-propagating plasma slab, which acquires the fraction $\approx 1-(4\gamma
)^{-2}$ of the driver pulse energy, \cite{bib:RPD}. The plasma slab also
acts as a mirror for a counter-propagating relativistically strong
electromagnetic radiation (the source). As such it exhibits the properties
of the sliding and oscillating mirrors, producing relativistic harmonics.
The source pulse should be sufficiently weaker than the driver, nevertheless
it can be relativistically strong. In the spectrum of the reflected
radiation, the fundamental frequency of the incident radiation and the
relativistic harmonics and other high-frequency radiation
like bremsstrahlung
generated at the
plasma slab are multiplied by the same factor, $\approx 4\gamma ^{2}$, Fig. 
\ref{fig:Kagami}.

Compared with previously discussed schemes,
the double-sided mirror concept
benefits from a high number of reflecting electrons
(since the accelerating plasma slab initially has solid density
and can be further compressed during the interaction)
and from the multiplication of the frequency of all the harmonics
(since the interaction is strongly nonlinear and the mirror is relativistic).
This concept opens the way towards extremely bright sources
of ultrashort energetic bursts of X- and gamma-ray, which become realizable
with present-day technology enabling new horizons of laboratory
astrophysics, laser-driven nuclear physics, and studying the fundamental
sciences, e.g. the nonlinear quantum electrodynamics effects.


\begin{figure}[Hb]
\includegraphics[scale=1.]{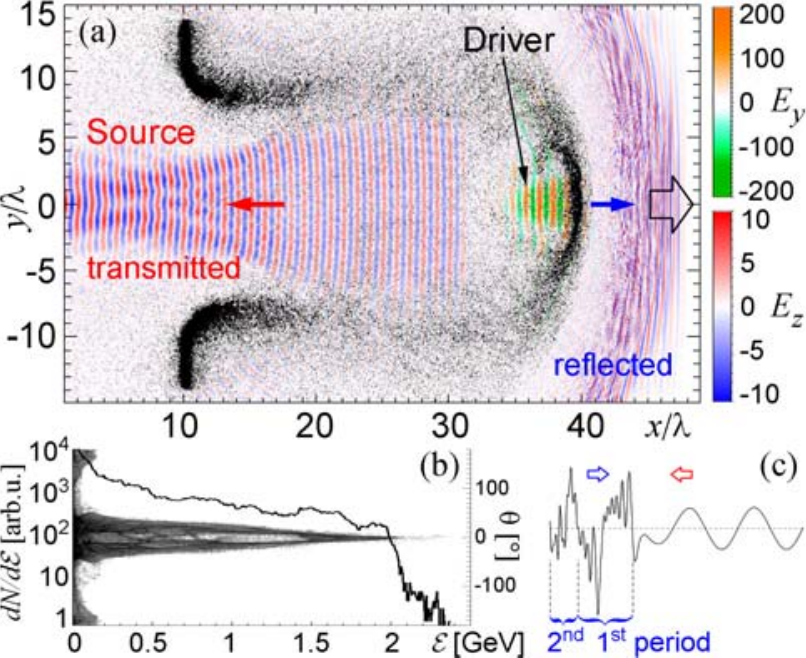}
\caption{(color)
(a) The driver and source pulses represented by the $y$- and $z$-components
of the electric field, respectively, and the ion density (black).
(b) The ion energy (curve) and anglular (grayscale) distributions.
Both the frames for $t=37\times 2\pi/\omega$.
(c) The electric field $z$-component, showing
the source pulse overlapped with the first two cycles
of the reflected radiation at $t=4\times 2\pi/\omega$.
\label{fig:2}
}
%
\includegraphics[scale=1.]{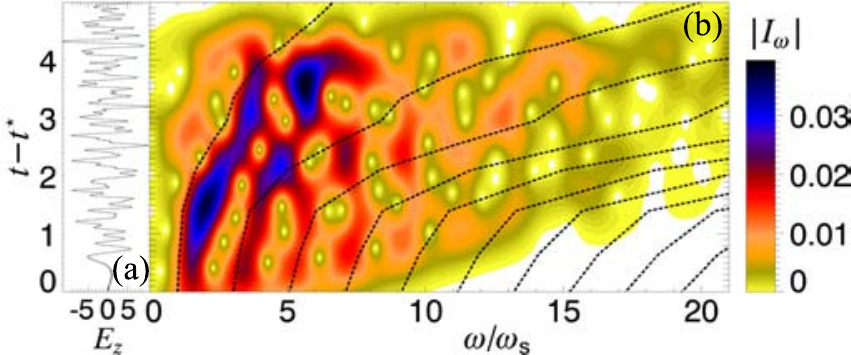}
\caption{(color)
(a) The electric field component $E_z$ representing the reflected radiation
along $x$-axis
for $t^\star=32\times 2\pi/\omega$.
(b) Colorscale: the modulus of the spectrum
of $E_z(\tau)$ seen along $x$-axis,
taken for each $t$ with the Gaussian filter,
$I_\omega(t)=\int_{-\infty}^{+infty}E_z(\tau)e^{-i\tau\omega-c^2(\tau-t)^2/\lambda^2}d\tau$.
Dashed curves: the odd harmonics frequency multiplied
by the factor $(1+\beta)/(1-\beta)$ calculated from the
fast ion spectrum maximum at the time of reflection.
Modes aliasing occurs at later times due to the fixed width of the filter and
a fast change of the frequency multiplication factor.
\label{fig:3}
}
\end{figure}
In order to investigate the feasibility of this effect we performed
two-dimensional (2D) particle-in-cell (PIC) simulations using the
Relativistic ElectroMagnetic Particle-mesh code REMP based on the density
decomposition scheme \cite{bib:REMP}. The driver laser pulse with the
wavelength $\lambda _{d}=\lambda=2\pi c/\omega $, the intensity $I_{d}=1.2\times 10^{23}$%
W/cm$^{2}\times (1\mu m/\lambda )^{2}$, corresponding to the dimensionless
amplitude $a_d=300$, and the duration $\tau_d =20\pi/\omega $ is focused with
the spot size of $D_{d}=10\lambda $ onto a hydrogen plasma slab with the
thickness $l=0.25\lambda $ and the initial electron density $%
n_{e}=480n_{cr}=5.4\times 10^{23}$cm$^{-3}\times (1\mu m/\lambda )^{2}$
placed at $x=10\lambda $.
The plasma slab transverse size is 28 $\lambda$.
The driver pulse shape is Gaussian but without
the leading part, starting $5\lambda $ from the pulse center along the $x$%
-axis. At the time $t=0$, when the driver pulse hits the plasma slab from the left (%
$x<10\lambda $), the source pulse arrives at another side of the slab from
the right ($x>10.25\lambda $). The driver is $p$-polarized, i.~e., its
electric field is directed along the $y$-axis. The source pulse is $s$%
-polarized (its electric field is along the $z$-axis). It has the same
wavelength as the driver pulse. Its intensity is $I_{s}=1.2\times 10^{19}$%
W/cm$^{2}\times (1\mu m/\lambda )^{2}$, corresponding to the dimensionless
amplitude $a_s=3$, its duration is $\tau_s =120\pi/\omega $ and its waist
size is $D_{s}=20\lambda $. The source pulse has rectangular profile along
the $x$- and $y$-axes; such the profile is not necessary for the desired
effect but helps to analyse the results.
The simulation box has size of 50 $\lambda$ with the resolution of
128 steps per $\lambda$ along the $x$-axis and 32 $\lambda$ with the
resolution 16 steps per $\lambda$ along the $y$-axis.  The number of
quasi-particles is $10^6$.
We note that the use of a
circularly polarized driver pulse  may provide a smoother start of the
slab acceleration in the radiation pressure dominant regime \cite
{bib:circ-RPDA}, nevertheless it was chosen to be $p$-polarized in order to
easily distiguish between the driver and the source pulses. In addition, our
choice demonstrates the robustness of the double-sided mirror effect. The
results are shown in Figs. \ref{fig:2}-\ref{fig:3}, where the spatial
coordinates and time units are in the laser wavelengths and wave periods,
respectively.

The driver laser makes a cocoon where it stays confined, Fig. \ref{fig:2}(a).
At $t=37\times 2\pi/\omega$, the ions are accelerated up to 2.4 GeV while
the majority of accelerated ions carry the energy about 1.5 GeV, Fig. \ref{fig:2}(b).
The accelerating plasma reflects the source pulse, which becomes
chirped and compressed about 10 times, Fig. \ref{fig:2}(a).
As the mirror velocity, $c\beta$, increases,
the reflected light frequency grows as $(1+\beta)/(1-\beta)$,
thus the electric field profile along the $x$-axis becomes more and more jagged, Fig. \ref{fig:3}(a).
A portion of the source
pulse reflected from the curved edges of the expanding cocoon acquires an
inhomogeneous frequency upshift determined by the angle of the reflecting
region. At the begining, the magnitude of the reflected radiation is higher
than that of the incident source ($\sim 3$ times), due to the
double Doppler effect and an enhancement of reflectivity
owing to the plasma slab compression under the radiation pressure exerted by the driver pulse.
%
In an instantaneous proper frame of the accelerating mirror, 
the frequency of the source pulse increases with time,
thus the mirror becomes more transparent.
Correspondingly, the source
starts to be transmitted through the plasma more efficiently, as
seen in Fig. \ref{fig:2}(a) where the transmitted radiation is focused
owing to a cocoon-like spatial distribution of the plasma.

The reflected radiation has a complex structure of the spectrum, Fig. \ref{fig:3}(b).
It contains not only the frequency-multiplied fundamental mode of the source
pulse, but also high harmonics due to the nonlinear interaction of the
source with the plasma slab.
This is also seen in Fig. \ref{fig:2}(c), where the first
two consecutive cycles of the reflected radiation exhibit
presence of high harmonics, while the later cycle is compressed together
with its harmonics in comparison with the earlier cycle.
The spectrum is also enriched by a continuous component
since the mirror moves with acceleration.
The high harmonics generation efficiency
is optimal for a certain areal density of the foil,
according to the condition $a_s\approx \pi n_e l r_e \lambda_s$ \cite{bib:FOIL},
where $r_{e}=e^{2}/m_{e}c^{2}$ is the classical electron radius.
Initially far from this condition, the accelerated plasma slab satisfies it
at certain time, when harmonics are generated most efficiently.


In order to analytically describe the reflected EM wave we use
the approximation of an infinitely thin foil (see also Refs.\cite{bib:FM,bib:FOIL}),
representing a mirror moving along the $x$-axis with the coordinate $\mathscr{X}_{M}(t)$.
We consider the one-dimensional (1D) Maxwell equation
\begin{equation}
\frac{\partial ^{2}\mathscr{A}}{\partial t^{2}}-{c^{2}}\frac{\partial ^{2}%
\mathscr{A}}{\partial x^{2}}+\frac{4\pi e^{2}n_{e}l\,\delta [x-\mathscr{X}%
_{M}(t)]}{m_{e}\gamma _{M}}\mathscr{A}=0\,,  \label{eq:1DMXWLL}
\end{equation}
where $\gamma _{M}=\left[ 1-(d\mathscr{X}_{M}/dt)^{2}c^{-2}\right] ^{-1/2}$
is the Lorentz factor of the mirror,
$\mathscr{A}$ is the EM wave vectror-potential,
$\delta$ is the Dirac delta function.
Let $k$ is the incident wave number.
Transfromations to dimensionless
variables and to new variables, $\xi ,\eta $, which are the characteristics
of the Maxwell equation,
\begin{eqnarray}
&
\bar{x} =kx\, ,\;\;\bar{t}=kct\, ,
&
\\
&
\xi =(\bar{x}-\bar{t})/2\, ,\;\;\eta
=(\bar{x}+\bar{t})/2\, ,  \label{eq:trans}
&
\\
&
\mathscr{X}_{M}(t) =k^{-1}X_{M}\left( {\eta -\xi }\right) \, ,\;\;%
\mathscr{A}(x,t)={\textstyle\frac{m_{e}c^{2}}{e}}A(\xi ,\eta )\, ,
&
\end{eqnarray}
and the property $\delta (kz)=k^{-1}\delta (z)$ yield the
equation
\begin{equation} \label{eq:axieta}
\frac{\partial ^{2}A}{\partial \xi \partial \eta}
=\chi \frac{\delta [\psi (\xi ,\eta)]}{\gamma (\xi ,\eta )} A,
\end{equation}
where $\chi =2n_{e}lr_{e}\lambda $, 
$\lambda =2\pi/k$, and
\begin{eqnarray}
\psi (\xi ,\eta )={\xi +\eta }-X_{M}\left( {\eta -\xi }\right) \, ,
\label{eq:PSIXIETA}
\\
\gamma (\xi ,\eta )=\left[ 1-X_{M}^{\prime 2}\left( {\eta -\xi }\right)
\right] ^{-1/2}\,.  \label{eq:GAM-XIETA}
\end{eqnarray}

We seek the solution to Eq. (\ref{eq:axieta}) in the form of the incident,
transmitted and reflected waves: 
\begin{equation}  \label{eq:asol}
A(\xi,\eta) = \left\{
\begin{array}{l}
a_1(\xi) + a_0 e^{2i\eta}, \quad \psi(\xi,\eta)>0 ; \\
a_2(\eta), \quad \psi(\xi,\eta)\le 0 .
\end{array}
\right.
\end{equation}
Here the factor $e^{2i\eta}=e^{ik(x+ct)}$ represents the incident wave. The
solution should satisfy the boundary conditions at the position of the
mirror, $\psi(\xi,\eta)=0$.
We introduce new functions $\xi_0(\eta)$ and $\eta_0(\xi)$,
which satisfy the following expressions
\begin{eqnarray}
\psi(\xi_0(\eta),\eta)=0 \hbox{  for  } \forall \eta
,
\\
\psi(\xi,\eta_0(\xi))=0 \hbox{  for  } \forall \xi
.
\end{eqnarray}
The requirement that
the solution is continuous, $A(\xi,\eta_0(\xi)-0) = A(\xi,\eta_0(\xi)+0)$,
leads to the following condition:
\begin{equation}  \label{eq:bc0}
a_1(\xi) + a_0 e^{2i\eta_0(\xi)} = a_2(\eta_0(\xi)) \, .
\end{equation}
The remaining conditions can be obtained from Eqs. (\ref{eq:axieta}) and (\ref{eq:asol}).
Integrating Eq. (\ref{eq:axieta}) over $\eta$
in the vicinity of $\eta_0(\xi)$ for fixed $\xi$
and some small $\epsilon>0$, we obtain:
\begin{equation}\label{eq:axi}
\left.
\frac{\partial A}{\partial\xi}
\right|_{\eta=\eta_0(\xi)-\epsilon}^{\eta=\eta_0(\xi)+\epsilon} =
\chi
\int_{\eta_0(\xi)-\epsilon}^{\eta_0(\xi)+\epsilon}
\delta[\psi(\xi,\eta)] \frac{A(\xi,\eta)}{\gamma(\xi,\eta)} d\eta
\, .
\end{equation}
If $X'_M\left(\eta_0(\xi)-\xi\right)\not=1$ for given $\xi$,
which means that the mirror velocity does not reach the speed of light in vacuum,
the function $\psi$ has a simple zero at $(\xi,\eta_0(\xi))$,
therefore the equation
\begin{equation}\label{eq:delta-id}
\int_{\eta_0-\epsilon}^{\eta_0+\epsilon}\delta[\psi(\xi,\eta)]f(\xi,\eta)d\eta =
\frac{f(\xi,\eta_0)}{\partial\psi/\partial\eta}
\,
,
\end{equation}
where the derivative $\partial\psi/\partial\eta$ is taken at the point
$\left(\xi,\eta_0(\xi)\right)$,
holds for any integrable function $f$.
Using Eq. (\ref{eq:delta-id}) we find that
at the limit $\epsilon\rightarrow 0$
Eq. (\ref{eq:axi}) gives the magnitude of the jump discontinuity
of the derivative $A_\xi=\partial A/\partial\xi$ at $\eta=\eta_0(\xi)$ for fixed $\xi$:
\begin{equation}\label{eq:axijump}
\left.
\frac{\partial A}{\partial\xi}
\right|_{\eta=\eta_0(\xi)-0}^{\eta=\eta_0(\xi)+0} =
\chi \digamma(\xi,\eta_0(\xi))
A(\xi,\eta_0(\xi))
\, ,
\end{equation}
where we introduce the digamma factor, $\digamma$,
\begin{equation}\label{eq:digamma}
\digamma(\xi,\eta) =
\left[
  \frac{1 + X'_M\left({\eta-\xi}\right)}{1 - X'_M\left({\eta-\xi}\right)}
  \right]^{1/2}
.
\end{equation}
Similar expression is obtained for the magnitude of the jump discontinuity
of the derivative $A_\eta=\partial A/\partial\eta$ at $\xi=\xi_0(\eta)$ for fixed $\eta$.
For the ansatz (\ref{eq:asol}) these expressions give the
following two ordinary differential equations (ODE):
\begin{eqnarray} \label{eq:bc1}
\lefteqn{}&&
a'_1(\xi) =
\chi
\!\left(  a_1(\xi) + a_0 e^{2i\eta_0(\xi)} \right)
\digamma(\xi,\eta_0(\xi))
\, ,
\\ \label{eq:bc2}
&&
2i a_0 e^{2i\eta} - a'_2(\eta) = 
\frac{\chi}{\digamma(\xi_0(\eta),\eta)} a_2(\eta)
\,
.
\end{eqnarray}
The reflected, $a_1(\xi)$, and the transmitted, $a_2(\eta)$, waves
are determined by
Eqs. (\ref{eq:bc0}), (\ref{eq:bc1}), and (\ref{eq:bc2})
which can be easily reduced to quadratures.

In the simplest case of uniform motion,
\begin{equation}
X^{\prime }_M(\bar t) = \beta=\mathrm{const},
\end{equation}
we have
\begin{eqnarray}
\digamma(\xi,\eta)= \digamma_0 = \left[\frac{1+\beta}{1-\beta}\right]^{1/2}\approx 2\gamma_M ,
\\
\eta_0(\xi) = -\digamma_0^2 \xi .
\end{eqnarray}
The solution to Eqs. (\ref{eq:bc0}), (\ref{eq:bc1}), and (\ref{eq:bc2}) reads
\begin{eqnarray}
a_1 = -a_0 \frac{\chi}{\chi+2i\digamma_0} \exp({-2i\digamma_0^2\xi}),
\\
a_2 = a_0\frac{2i\digamma_0}{\chi+2i\digamma_0} \exp({2i\eta}),
\end{eqnarray}
so that the reflection coefficient in terms
of the number of photons is
\begin{equation}
\mathsf{R}=\left|\frac{a_1}{a_0}\right|^2 \approx
\frac{(n_e l r_e \lambda)^2}{(n_e l r_e \lambda)^2+4\gamma_M^2}
,
\end{equation}
thus we recover the result
of Ref. \cite{bib:FM}.

In the case of a mirror moving with a uniform acceleration $g k c^2$, for
simplicity we consider the particular trajectory
\begin{equation}
X_M(\bar t) = g^{-1}[1+(g\bar t)^2 ]^{1/2}.
\end{equation}
Then we obtain
\begin{eqnarray}
&
\eta_0(\xi)=(4g^2\xi)^{-1}, \;\;
\xi_0(\eta)=(4g^2\eta)^{-1}
&
\\
&
\digamma(\xi,\eta_0(\xi)) = (2g\xi)^{-1}, \;\;
\digamma(\xi_0(\eta),\eta) = 2g\eta,
&
\end{eqnarray}
and the solution to Eqs. (\ref{eq:bc0}), (\ref{eq:bc1}), (\ref
{eq:bc2}):
\begin{eqnarray}  \label{eq:a2eta}
a_1(\xi) = {\textstyle\frac{\chi a_0}{2g}} \left(2i g^2\xi\right)^{\frac{\chi}{2g}}
\Gamma\left[{\textstyle\frac{\chi}{2g}},
({2ig^2\xi})^{-1},0\right] , \\
a_2(\eta) = {\textstyle\frac{\chi a_0}{2g}} \left(-2i\eta\right)^{-\frac{\chi}{2g}}
\Gamma\left[{\textstyle\frac{\chi}{2g}},-2i\eta,0\right]+a_0 e^{2i\eta} ,
\end{eqnarray}
where $\Gamma(a,z_1,z_2)=\int_{z_1}^{z_2}t^{a-1}e^{-t}dt$ is the generalized
incomplete gamma function \cite{bib:MathBook}.
At $\xi \rightarrow 0$,
\begin{eqnarray}
a_1(\xi) = -{\textstyle\frac{\chi a_0}{2g}}
	\left(2i g^2\xi\right)^{\frac{\chi}{2g}} \Gamma\left({\textstyle\frac{\chi}{2g}}\right)
	+{i \chi a_0 g}\exp\left({\frac{i}{2g^2\xi}}\right)
	\left(\xi+O(\xi^2)\right) ,
\end{eqnarray}
where $\Gamma(z)$ is the Euler gamma function \cite{bib:MathBook}. The frequency
of the reflected radiation increases as $\xi^{-1}$, as in the case of a
perfect mirror of Ref. \cite{bib:Hartemann}. However, in our case the mirror
reflectivity decreases with time.
An observer at infinity (which corresponds to $\xi=0$)
see the radiation with the frequency spectrum of the following intensity
\begin{equation}
I_\nu \approx \frac{\pi a_0^2}{2g^4(1+2g/\chi)^2}
\left[
{_1{\rm F}_2}\left( 1+{\textstyle\frac{\chi}{2g}};2,2+{\textstyle\frac{\chi}{2g}};-\frac{\nu}{2g^2} \right)
\right]^2
,
\end{equation}
where $\nu$ is the observed frequency,
${_1{\rm F}_2}(a_1;b_1,b_2;z)$
is the generalized hypergeometric function \cite{bib:MathBook}.
Here $I_\nu$ is defined as the square of the modulus of the Fourier transform of
the function $a_1(\xi)$,
$I_\nu = \left|\frac{1}{\sqrt{2\pi}}\int_{-\infty}^{+\infty}e^{i\nu\xi}a_1(\xi)d\xi\right|^2$,
where we cast out the essential singularity at $\nu=0$ representing the finite limit
$a_1(\xi\rightarrow+\infty)=-a_0$.
For large $\nu$, the spectral intensity decreases with frequency, $\nu$, as
\begin{equation}
I_\nu \sim
\frac{a_0^2}{4g^4}
\left\{
\frac{\sqrt{2\pi} \Gamma(1+{\textstyle\frac{\chi}{2g}})}{\Gamma(-{\textstyle\frac{\chi}{2g}})}
\left(\frac{2g^2}{\nu}\right)^{1+\frac{\chi}{2g}}
+
\frac{\chi}{2g}
\left[
\cos\left({\textstyle\frac{\sqrt{2\nu}}{g}}\right)
+
\sin\left({\textstyle\frac{\sqrt{2\nu}}{g}}\right)
\right]
\left(\frac{2g^2}{\nu}\right)^{5/4}
\right\}^2
,\;\; \nu\rightarrow\infty
.
\end{equation}

Now we consider the case of a mirror oscillating with frequency $\Omega$
(normalized on the incident wave frequency)
\begin{equation}  \label{eq:Vbart}
\frac{d}{d\bar t}\left(\frac{\beta(\bar t)}{\sqrt{1-\beta^2(\bar t)}}\right) =g \cos(\Omega
\bar t) \, ,
\end{equation}
choosing the following trajectory of the mirror
\begin{equation}  \label{eq:XMbart}
X_M(\bar t) =
\frac{1}{\Omega}
\arctan\left(
-\frac{\cos(\Omega\bar t)}{\sqrt{h^2 - \cos^2(\Omega\bar t)}}
\right)
\, .
\end{equation}
where $h^2 = 1+\Omega^2/g^2$.
Solving the equation $\phi(\xi,\eta_0(\xi))=0$, we obtain
\begin{eqnarray}\label{eq:}
\eta_0(\xi)=
-\frac{1}{\Omega}
\arctan H
-\frac{\pi}{\Omega} \left\lfloor \frac{\Omega\xi + \arctan h}{\pi} - \frac{1}{2} \right\rceil
,
\\
H=\frac{h\tan(\Omega\xi)+1}{\tan(\Omega\xi)+h},
\end{eqnarray}
where the function $\lfloor z\rceil$ gives the integer closest to $z$.
The digamma factor, Eq. (\ref{eq:digamma}), for $\eta=\eta_0(\xi)$ reads
\begin{equation}  \label{eq:digam}
\digamma(\xi,\eta_0(\xi)) = \left[\frac{h^2-1}{h^2+1+2h\sin(2\Omega\xi)}\right]^{1/2} \, .
\end{equation}
For Eq. (\ref{eq:XMbart}) the only bounded solution to Eq. (\ref{eq:bc1}) is
\begin{eqnarray}  \label{eq:a1xi1}
a_1(\xi) = \frac{\chi a_0}{g} \int\limits_{\Omega\xi}^{+\infty} \frac{%
E(\Omega\xi)}{E(\tau)} \frac{e^{-\frac{2i\tau}{\Omega}} (h-ie^{2i\tau})^{\frac{2}{\Omega}}
\, \, d\tau} {\left(h^2+1+2h\sin(2\tau)\right)^{\frac{2+\Omega}{2\Omega}}} , \\
E(\tau) = \exp\left\{ {\textstyle\frac{\chi}{g(h+1)}} \mathrm{F}\left(\left. \tau -
{\textstyle\frac{\pi}{4}} \right| {\textstyle\frac{4h}{(h+1)^2}} \right) \right\} \, .
\end{eqnarray}
where $\mathrm{F}(z|m)$ 
is the elliptic integral of the first kind with an asymptotic $\propto z$
for $z\rightarrow\infty$ \cite{bib:MathBook}.


In conclusion, a solid density plasma slab, accelerated in the radiation
pressure dominant regime, can efficiently reflect a counter-propagating
relativistically strong laser pulse (source). The reflected electromagnetic
radiation consists of the reflected fundamental mode and high harmonics, all
multiplied by the factor $\approx 4\gamma ^{2}$, where $\gamma $ increases
with time. In general, the reflected radiation is chirped due to the mirror
acceleration. With a sufficiently short source pulse being sent with an
appropriate delay to the accelerating mirror, one can obtain a high-intense
ultra-short pulse of X-rays.

For the mirror velocities greater than some threshold, the distance between
electrons in the plasma slab in the proper reference frame becomes longer
than the incident wavelength. Thus the plasma slab will not be able to
afford the reflection in a coherent manner, where the reflected radiation
power is proportional to the square of the number of particles in the
mirror. In this case the reflected radiation becomes linearly proportional
to the number of particles. Even with this scaling one can build an
ultra-high power source of short gamma-ray pulses, when the interaction of
the source pulse with a solid-density plasma is in the regime of the
backward Thomson scattering.

We can estimate the reflected radiation brightness in two limiting cases.
For $2\gamma < (n_e\lambda_s^3)^{1/6}$,
the reflection is coherent and
for the reflected photon energy $\hbar\omega$
the brightness is $B_M\approx {\cal E}_s(\hbar\omega)^3\lambda_s/4\pi^5 \hbar^4 c^3$,
where ${\cal E}_s$ is the source pulse energy.
For larger $\gamma$,
the interaction may become incoherent and should be described as Thomson scattering,
which gives $B_T \approx a_d{\cal E}_s(\hbar\omega)^2r_e\lambda_s^2/8\pi^4 \hbar^3 c^2\lambda_d^3$.
For ${\cal E}_s = 10$J, $\lambda_s = 0.8 \mu$m, $\hbar\omega = 1$keV ($\gamma =13$),
$B_M = 0.8\times 10^{40}$photons/mm$^2$mrad$^2$s,
which is orders of magnitude greater than any existing or proposed source, \cite{bib:SOURCES}.
For the same parameters and $\lambda_d = 0.8\mu$m,
$a_d = 300$, $\hbar\omega = 10$keV ($\gamma =40$),
$B_T = 3\times 10^{32}$photons/mm$^2$mrad$^2$s.

Employing the concept of the accelerating double-sided mirror, one can
develop a relatively compact and tunable ultra-bright high-power X-ray or
gamma-ray source, which will considerably expand the range of applications
of the present-day powerful sources and will create new applications and
research fields. Implementation in the ``water window'' will allow performing
a single shot high contrast imaging of biological objects \cite{bib:Haj}. In atomic physics
and spectroscopy, it will allow performing the multi-photon ionization and
producing high-Z hollow atoms. In material sciences, it will reveal novel
properties of matter exposed to the high power X-rays and gamma-rays. In
nuclear physics it will allow studying states of high-Z nucleus. The sources
of high-power coherent X-ray and ultra-bright gamma-ray radiation also pave
the way towards inducing and probing the nonlinear quantum
electrodynamics processes.


This research was partially supported by the
Ministry of Education, Science, Sports and Culture,
Grant-in-Aid for Scientific Research (A), 29244065, 2008.

\end{document}